\documentclass{article}
\pdfoutput=1
\usepackage{PRIMEarxiv}

\usepackage[utf8]{inputenc} % allow utf-8 input
\usepackage[T1]{fontenc}    % use 8-bit T1 fonts
\usepackage{url}            % simple URL typesetting
\usepackage{booktabs}       % professional-quality tables
\usepackage{nicefrac}       % compact symbols for 1/2, etc.
\usepackage{microtype}      % microtypography
\usepackage{lipsum}
\usepackage{fancyhdr}       % header
\usepackage{graphicx}       % graphics
\graphicspath{{media/}}     % organize your images and other figures under media/ folder
\usepackage[colorlinks = true,
            linkcolor = blue,
            urlcolor  = blue,
            citecolor = blue,
            anchorcolor = blue]{hyperref}
\usepackage{makecell}
\usepackage{array}
\usepackage{enumitem}
\usepackage{amsmath,amssymb,amsfonts}
\usepackage{gensymb}

\title{open-UST: An Open-Source Ultrasound Tomography Transducer Array System \thanks{Paper submitted for review on 13/02/2023. This work was supported by the National Physical Laboratory, European Union’s Horizon 2020 research and innovation programme H2020 ICT 2016-2017 (Grant No. 732411, as an initiative of the Photonics Public Private Partnership), the Engineering and Physical Sciences Research Council (EPSRC), UK, (Grant Nos. EP/S026371/1, EP/T014369/1 and EP/P008860/1), a UKRI Future Leaders Fellowship (Grant No. MR/T019166/1), and the EPSRC Centre For Doctoral Training in Intelligent Integrated Imaging In Healthcare (i4health). For the purpose of open access, the author has applied a Creative Commons Attribution (CC BY) license to any Author Accepted Manuscript version arising (corresponding author: Morgan Roberts). Eleanor Martin is also with the Wellcome/EPSRC Centre for Interventional and Surgical Sciences, University College London, London, UK.}
}
\author{
  Morgan Roberts, Eleanor Martin, Michael D. Brown, Ben T. Cox, Bradley E. Treeby \\
  Department of Medical Physics and Biomedical Engineering \\
  University College London, Gower Street, London, WC1E 6BT, UK \\
  \texttt{morgan.roberts.18@ucl.ac.uk} \\}

\begin{document}
\maketitle

\begin{abstract}
Fast imaging methods are needed to promote widespread clinical adoption of Ultrasound Tomography (UST), and more widely available UST hardware could support the experimental validation of new measurement configurations. In this work, an open-source 256-element transducer ring array was developed (\href{https://morganjroberts.github.io/open-UST/}{morganjroberts.github.io/open-UST}) and manufactured using rapid prototyping, for only £2k. Novel manufacturing techniques were used, resulting in a 1.17$\degree$ mean beam axis skew angle, a 104 $\mu$m mean element position error, and a $\pm$13.6 $\mu$m deviation in matching layer thickness. The nominal acoustic performance was measured using hydrophone scans and watershot data, and the 61.2 dB SNR, 55.4$\degree$ opening angle, 16.3 mm beamwidth and 54\% transmit-receive bandwidth (-12 dB), were found to be similar to existing systems, and compatible with full waveform inversion reconstruction methods. The inter-element variation in acoustic performance was typically $<$10\% without using normalisation, meaning that the elements can be modelled identically during image reconstruction, removing the need for individual source definitions based on hydrophone measurements. Finally, data from a phantom experiment was successfully reconstructed. These results demonstrate that the open-UST system is accessible for users, and suitable for UST imaging research.
\end{abstract}

\section{Introduction} \label{sec:introduction}
Breast cancer screening reduces mortality, but mammograms have lower sensitivity for people with high breast density, and over-diagnosis causes harm in healthy people \cite{marmot2013benefits}. Ultrasound Tomography (UST) is a method for measuring the 3D acoustic property distributions in the breast, using a transducer array to transmit ultrasound waves into the breast from different angles and measure the transmitted and scattered field. Clinical performance has been promising \cite{wiskin2020full, duric2020using}, and advanced methods are being developed \cite{lucka2021high, bachmann2020source}, but improvement is still needed to achieve clinically useful image reconstruction times. UST hardware allows new measurement configurations to be investigated, but there is a high barrier to entry since UST systems are not available off the shelf, and custom arrays are expensive. Therefore, more widely available and re-configurable UST hardware could accelerate progress towards fast, accurate imaging methods and promote widespread clinical adoption of UST.

\begin{figure}[!t]
\centerline{\includegraphics[width=0.5\linewidth]{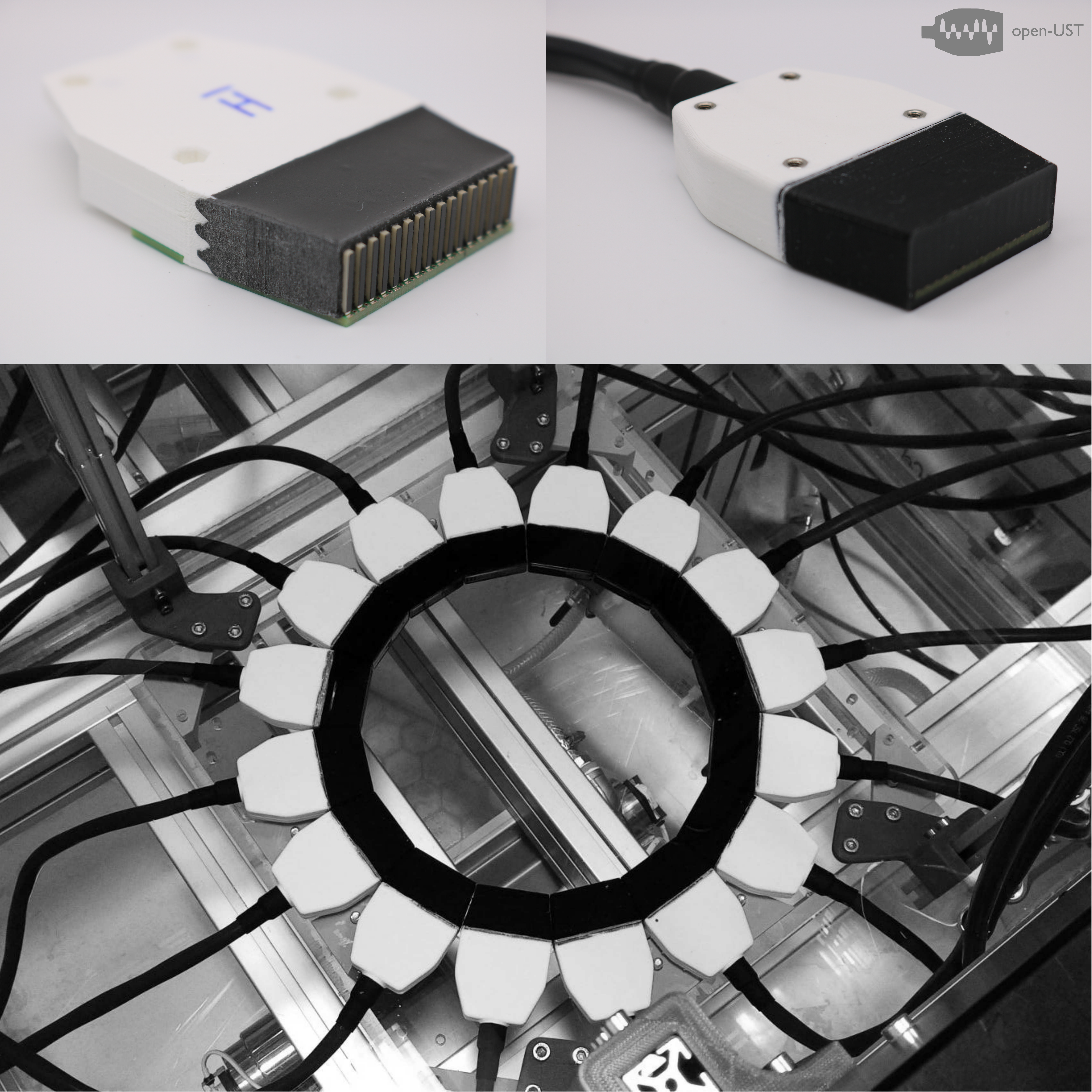}}
\caption{\textbf{Top Left:} Transducer during manufacture. \textbf{Top Right:} Finished transducer. \textbf{Bottom:} open-UST transducer array.}
\label{fig:USTarray_module_logo}
\end{figure}

Rapid prototyping technologies such as 3D-printing can be used to manufacture ultrasound hardware in-house, without expensive specialist equipment \cite{kim2014rapid, yddal2015open}. Rapid prototyped ultrasound hardware is low cost, has a short lead time, and can be easily modified, but the user has to design components from scratch. However, rapid prototyped hardware can also be easily open-sourced, a concept that has already promoted collaboration in the UST community \cite{cueto2022stride, ruiter2018usct}, which reduces the upfront design time, and could allow users without transducer manufacture experience to build a UST system in-house.

Open-source designs have been released for a microbubble characterisation chamber \cite{smith2022open} and an acoustic levitation system \cite{marzo2017ultraino}, and the files were accompanied by manufacturing instructions, which are essential for users to access the project. However, an open-source instructional guide for ultrasound transducer array manufacture does not exist. This paper presents the design, manufacture and evaluation of \href{https://morganjroberts.github.io/open-UST/}{open-UST}: a low cost UST transducer array system optimised for in-house manufacture using rapid prototyping (Figure \ref{fig:USTarray_module_logo}) \cite{openUST}. The hardware distribution includes CAD models, PCB and 3D-printing files, a bill of materials, assembly videos, and full manufacture documentation. End users are expected to be UST researchers comfortable with manual assembly processes, for example soldering and polymer casting. The goals of the open-UST project, explained below, are that:
\begin{enumerate}[label=(\Alph*)]
    \item The manufacture is accessible to users,
    \item The design parameters and nominal acoustic behaviour are suitable for UST imaging research,
    \item The inter-element variation (IEV) in acoustic behaviour is low.
\end{enumerate}

\subsection{Accessibility}
For users to access open-UST, the cost and lead time of the system should be low, meaning that the material cost should be minimised, and that only essential features should be included to accelerate assembly. Also, the manufacturing processes should be simple, and it is assumed that users do not have access to specialist transducer manufacture equipment, only a vacuum chamber, a 3D-printer capable of printing polylactic acid (PLA) and polyvinyl alcohol (PVA) filament, and standard workshop hand tools.

\subsection{Design Parameters and Nominal Acoustic Behaviour}
The bandwidth and signal-to-noise-ratio (SNR) of the open-UST transducers should provide useful data at frequencies above 350 kHz \cite{wiskin2020full}, to be compatible with full waveform inversion (FWI) reconstruction methods. The finite bandwidth and size of physical transducer elements affect the way that they emit and respond to ultrasound waves, and modelling their angle dependent frequency response (ADR) during image reconstruction can help to better match simulated and observed data, leading to increased reconstruction accuracy \cite{cueto2021spatial}. The open-UST transducers should have a smooth ADR, since this makes the UST data easier to interpret, and allows the response to be easily incorporated into the reconstruction forward model, for example by representing the transducers as ideal pistons with an effective element size chosen to best represent the watershot data. 

\subsection{Interelement Variation (IEV) in Acoustic Behaviour}
For transducer arrays, manufacturing tolerances can cause the individual elements to have slightly different ADRs, small variations in position and beam axis skews. These can be characterised to improve reconstruction accuracy \cite{cueto2021spatial}, but this requires additional time, hydrophone measurements, and computational complexity, presenting another barrier to entry to users. Although reconstruction methods exist that have achieved excellent results without modelling the ADR of the transducer elements \cite{bachmann2020source}, their implicit assumption is still that the transducers behave identically. Therefore, the IEV in acoustic behaviour of the open-UST system must be low so that the transducers can be modelled identically. 

For high performance arrays, low IEV is achieved using high precision manufacturing equipment, for example dicing saws, spin coaters and lapping machines, and so a tradeoff between cost and IEV is expected for rapid prototyped transducer arrays. Previously, the IEV in electrical impedance was measured for a 3D-printed histotripsy array \cite{kim2014rapid}. In this paper, the IEV in electrical impedance, transmit impulse response, beam axis skew, beamwidth, opening angle, signal-to-noise-ratio, receive crosstalk and transmit-receive directional response are assessed for the open-UST system.

Previously, prototype transducer modules were evaluated for open-UST \cite{roberts2021}, and low cost techniques for matching layer deposition were developed \cite{roberts2022matching}. In this paper, the design and manufacture of the open-UST system in explained, and then the experimental evaluation of IEV in acoustic performance is described. Finally, results from a phantom imaging experiment are shown as a proof of principle of the open-UST system.

\section{Design and Manufacture} \label{sect: design}

\subsection{Array Design}
The open-UST aperture configuration and acoustic performance should support typical UST imaging use cases, and facilitate experimentation with new arrangements. Two single-element or clinical array transducers could be purchased and mounted to a rotation stage to sample a virtual array \cite{gemmeke20073d, camacho2018multimodal}, but this configuration has a large data acquisition time and so multi-element transducer arrays that fully surround the object are typically used instead. Either bowl \cite{zapf2022realization} or rotating planar \cite{malik2018quantitative} configurations are used in 3D, but the most common design is a vertically translated 2D ring array \cite{roy2013breast, camacho2018multimodal, CUDEIROBLANCO20221995}, since these allow data to be collected and reconstructed in 2D slices, which is computationally efficient. The standard open-UST configuration is a 2D ring array, but its modularity also allows reconfiguration into 3D geometries. To simplify manufacture, each module is a linear array, with a total of 16 modules forming a hexadecagon approximation to a ring.

Current UST systems have between 40 \cite{CUDEIROBLANCO20221995} and 2304 \cite{zapf2022realization} transducer elements. Systems with many elements have denser sampling and higher image quality, but are complex to manufacture and require a data acquisition system (DAQ) with an equivalent channel count, or a multiplexer. Rotating the array can increase the sampling density, but adds complexity and cost, and increases the data acquisition time. The standard open-UST configuration is a 256-element ring array, since this is a typical number of channels available from open ultrasound DAQ platforms \cite{boni2018ultrasound}. Excellent reconstructions of in vivo data from 256-element ring arrays have been demonstrated using full waveform inversion (FWI) reconstruction methods \cite{agudo20183d}. 

The open-UST array diameter is 220 mm, which is larger than the pendant breast diameter for an entire study population of American women \cite{huang2011characterization}. The diameter and number of elements constrains the intra-module element pitch, which was chosen to be 2.54 mm to align with common PCB connector sizes. An overview of the array design is shown in Table \ref{table: parameters}.

\begin{table}[h] 
\centering
\caption{Key parameters of the open-UST transducer array.}
\begin{tabular}[t]{>{\raggedright}p{0.3\columnwidth}>{\raggedright\arraybackslash}p{0.15\columnwidth}} 
\toprule
Parameter & Value \\
\midrule
Number of elements & 256 \\
Number of transducer modules&16\\
Elements per module&16\\
PZT element thickness&1 mm\\
PZT element width&1 mm\\
PZT element length&10 mm\\
PZT pitch&2.54 mm\\
Material costs&£2k\\
Assembly time&4 months\\
\bottomrule
\label{table: parameters}
\end{tabular}
\end{table}

\subsection{PZT Element Selection} \label{sect: pzt_selection}
The open-UST system uses individual PZT plates for the transducer elements, since users are unlikely to have access to a dicing saw, and custom diced PZT slabs are expensive. PZT 850 was selected as a piezoelectric material, which is ideal for sensing applications, with an acoustic impedance of $Z_p$ = 31.5 MRayl, and frequency constants of $N_\text{T}$ = 2040 m/s and $N_\text{L}$ = 1500 m/s in the thickness and lateral directions respectively \cite{apc_materials}. The dimensions of the PZT elements affect their resonance spectra and beam patterns. For breast UST, the centre frequency is typically between 0.9 MHz \cite{malik2018quantitative} and 3 MHz \cite{roy2013breast}, and for 2D ring arrays a wide lateral opening angle and thin elevation beamwidth are required to confine the waves to a slice through the entire breast.

\begin{figure}[!h]
\centerline{\includegraphics[width=0.6\linewidth]{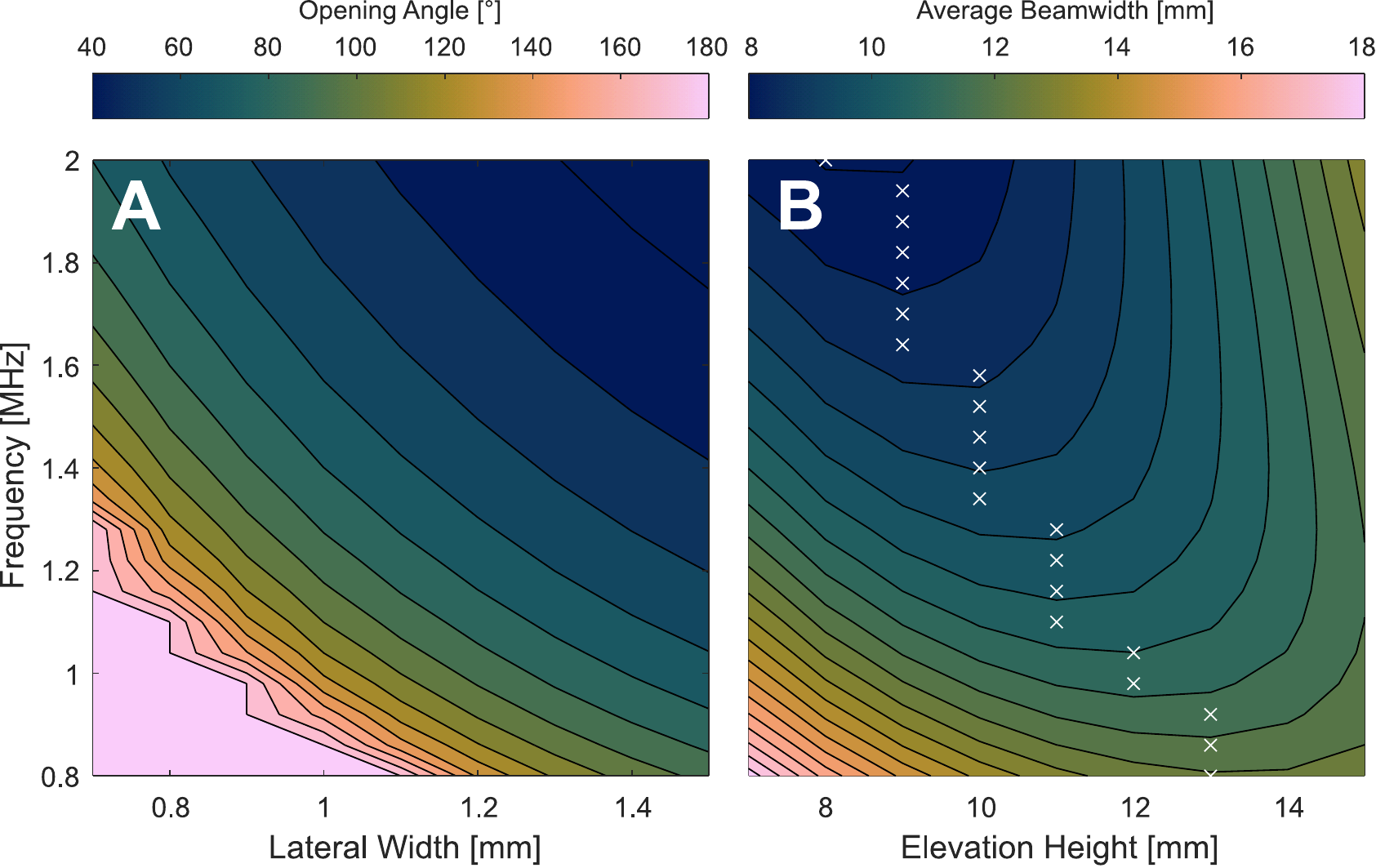}}
\caption{\textbf{A:} Predicted opening angle as a function of source width and frequency. \textbf{B:} Simulated average beam width as a function of source height and frequency. White crosses indicate the optimal elevation height with minimum beamwidth at each frequency.} 
\label{fig:beamwith_opening_angle_AFP_simulations}
\end{figure}

Decreasing the lateral width of PZT elements increases their lateral opening angle, but decreases their sensitivity and presents manufacturing challenges. The minimum PZT plate width widely available off-the-shelf is 1 mm. Figure \ref{fig:beamwith_opening_angle_AFP_simulations} shows opening angle predictions for lateral widths from 0.7 mm to 1.5 mm between 800 kHz and 2 MHz, simulated using the acoustic field propagator \cite{treeby2018rapid} from the k-Wave toolbox. For each simulation, the -6 dB opening angle was extracted from the far field directional response in the lateral plane. A width of 1 mm provides a minimum opening angle of 54$\degree$ at 2 MHz, which is comparable to the 43$\degree$ -10 dB opening angle at 2.6 MHz achieved by other UST systems \cite{angerer2022single}, indicating that this is a suitable beam pattern for UST imaging.

A 1 mm PZT thickness was selected due to its availability. A width to thickness aspect ratio $w/t\approx$ 1 can cause complex behaviour due to the interaction of lateral and thickness vibration modes \cite{de1985vibration}, meaning the exact plate resonances could not be calculated from the width and thickness. However, approximate thickness and lateral resonances of 2 MHz and 1.5 MHz were predicted, which are within the required range for UST.

Lenses can provide a thin and uniform elevational beamwidth \cite{sandhu20173d}, but these add complexity to the manufacturing procedure, and instead the elevation height of the elements was optimised to provide weak focusing. Figure \ref{fig:beamwith_opening_angle_AFP_simulations} shows -6 dB elevational beamwidth predictions, averaged over all axial positions from the source to the array radius, for elevational heights from 7 mm to 15 mm, simulated using the acoustic field propagator. The elevation height was selected as 10 mm due to its off the shelf availability, which is the optimal value from 1.34 MHz to 1.58 MHz. Following the dimension selection (summarised in Table \ref{table: parameters}), 1 $\times$ 1 $\times$ 10 mm PZT elements (Item 689, APC International Ltd, PA, USA) were purchased.

\begin{figure}[!h]
\centerline{\includegraphics[width=0.5\linewidth]{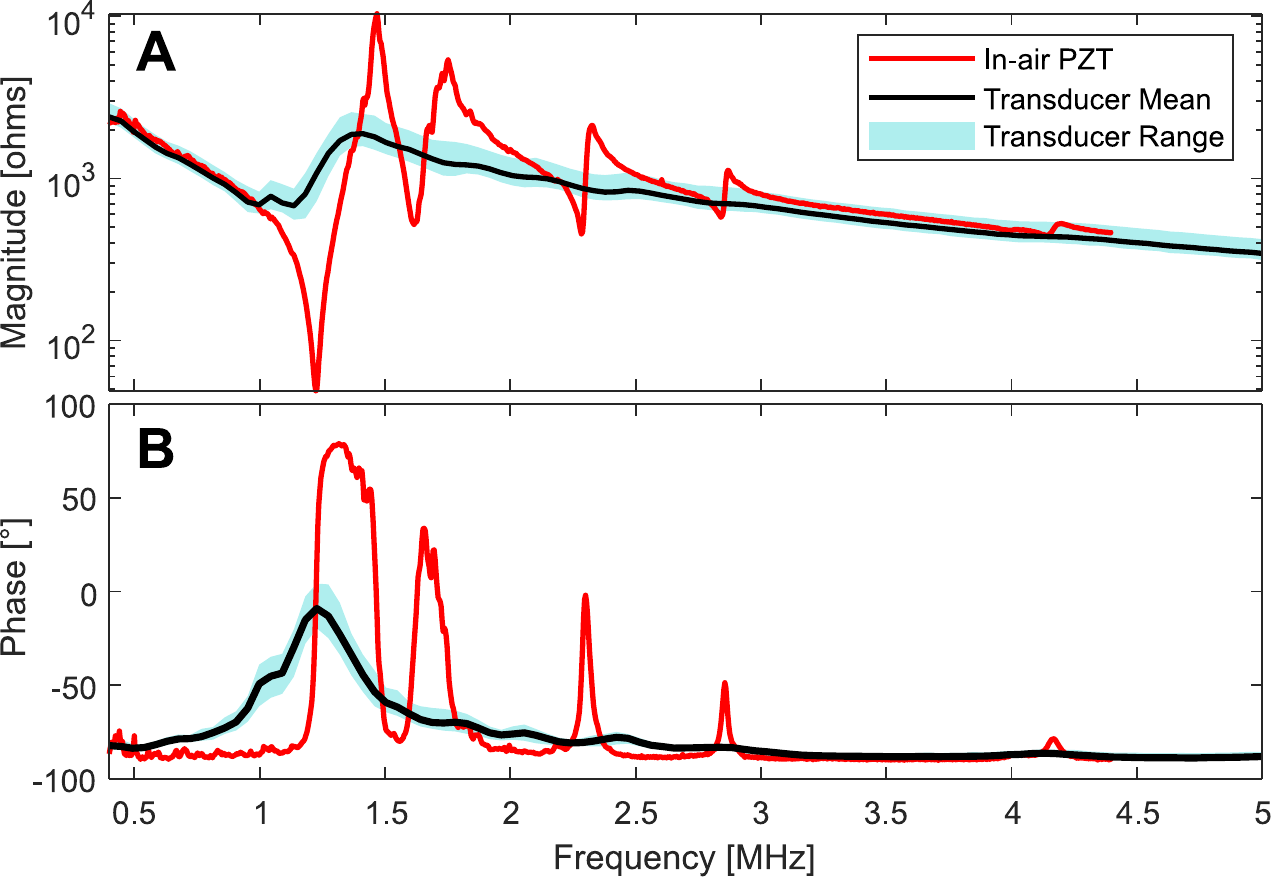}}
\caption{Electrical impedance magnitude (\textbf{A}) and phase (\textbf{B}) spectra of 128 transducer elements measured in water, after backing layer casting. The mean and entire measured range of the data are shown. Red lines show the in-air electrical input impedance spectra for a single PZT plate before manufacture.}
\label{fig: electrical_impedance}
\end{figure}

Figure \ref{fig: electrical_impedance} shows the in-air electrical input impedance spectrum of a single PZT plate, measured using a vector impedance analyser (4193A, Hewlett Packard). There are multiple resonances in the phase spectrum from 1.22 MHz to 2.86 MHz, which is a wider range than originally predicted, highlighting the difficulty in estimating resonances from the element dimensions alone. This spectrum suggested that the acoustic centre frequency would be close to 1.22 MHz, due to the strong series resonance caused by the lateral vibration mode coupled into the thickness mode. With a damping backing layer, the acoustic response was expected to tail off smoothly towards 2.86 MHz, providing a suitable bandwidth for UST imaging. Although the 10 mm elevation height is optimal for 1.34 MHz - 1.58 MHz, at the 1.22 MHz resonance, the average beamwidth was 10.24 mm, which is only slightly larger than the optimal value of 10.15 mm.

\subsection{Acoustic Stack Design and Manufacture } \label{sect: acoustic_stack}
\begin{figure}[!t]
\centerline{\includegraphics[width=0.49\columnwidth]{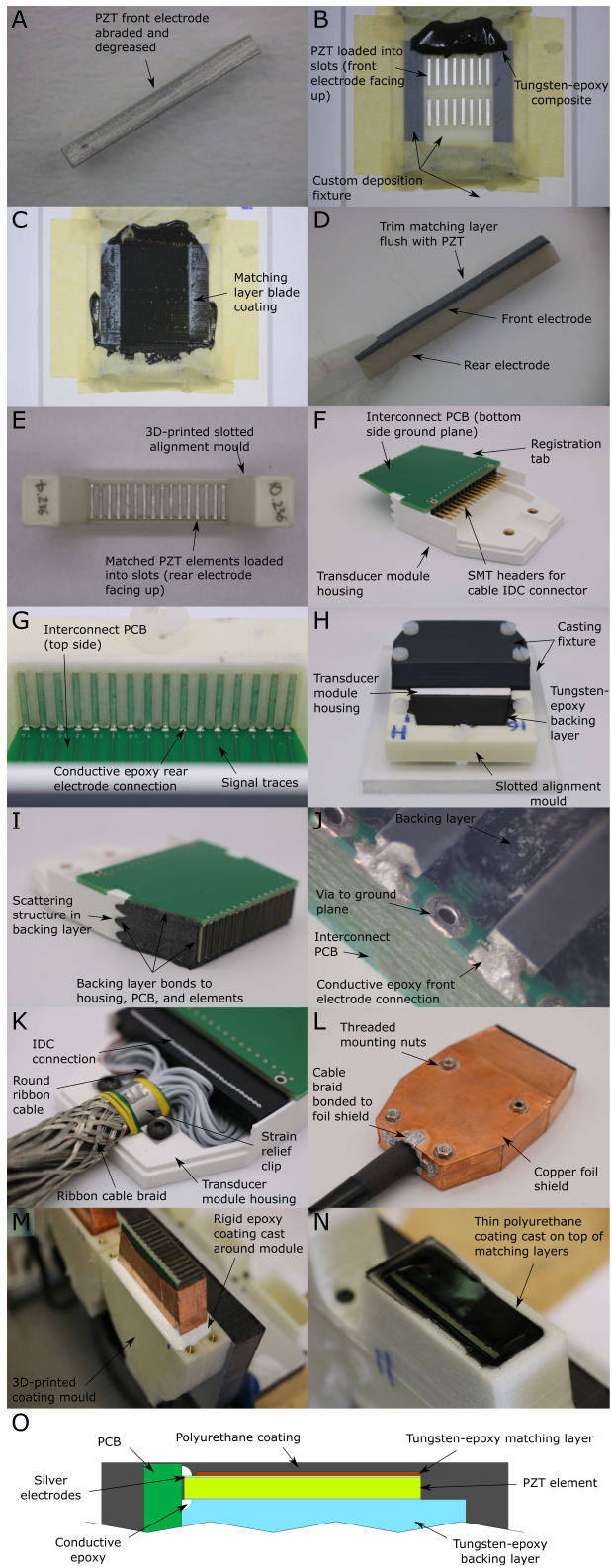}}
\caption{\textbf{A-N}: Summary of the transducer module manufacture. \textbf{O}: Cross section through the acoustic stack.}
\label{fig:assembly_summary_gallery}
\end{figure}

The acoustic stack for each element, shown in Figure \ref{fig:assembly_summary_gallery}O, is a PZT plate with a backing layer, quarter wavelength matching layer, and a thin polyurethane waterproof coating. Although formulations exist that account for complex resonant behaviour, acoustic stack design tools such as the KLM model were not used, since these rely heavily on accurate material properties, which are usually found by fitting to experimental data \cite{angerer2020enhanced}, and this was not available. The transducer module manufacturing procedure is documented on the \href{https://morganjroberts.github.io/open-UST/}{open-UST} website \cite{openUST}, and is summarised in Figure \ref{fig:assembly_summary_gallery}.

Matching and backing layer composites can be made by mixing filler powder with castable polymers. Tungsten powder was chosen because relatively low volume fractions can produce high impedance composites, which results in a low enough viscosity for the composite to be properly hand mixed. Araldite Standard epoxy (Huntsman Advanced Materials, Cambridge, UK) was chosen for the polymer since it is widely available, low cost, and has a high enough viscosity to prevent particle settling during curing.

Matching layers are typically tuned to the existing PZT resonance to achieve an even larger response. However, prototyping showed that the transmit pressure at 1.22 MHz was sufficient without a matching layer (Figure \ref{fig:transmit_impulse_response_probes_AEFG}). Instead, the matching layer resonance frequency was selected to be 2 MHz, which is higher than the main resonance and in the middle of the 1.22 MHz - 2.86 MHz range where an acoustic response was expected after damping. That was done because boosting the high frequency content is useful for improving resolution during image reconstruction. Figure \ref{fig:transmit_impulse_response_probes_AEFG} shows that this design worked as intended.
For a tungsten-epoxy composite used to match PZT with impedance $Z_p$ = 31.5 MRayl to water with impedance $Z_w$ = 1.5 MRayl, the target impedance of the matching layer should be $Z_l = \sqrt{Z_pZ_w}$ = 6.87 MRayl. Preliminary testing showed that a tungsten weight fraction of 86.7\% provides a sound speed of 1317 m/s, an acoustic impedance of 6.67 MRayl \cite{roberts2019}, and requires a quarter-wavelength matching layer thickness of 165 $\mu$m. For manufacture, a low cost deposition method was developed, first using blade coating (Figure \ref{fig:assembly_summary_gallery}B-D), and then compression between glass plates \cite{roberts2022matching}, producing a thickness distribution of 174 $\mu$m $\pm$ 13.6 $\mu$m (N = 128).

A backing layer was added to the rear face of each PZT element to increase damping and widen the bandwidth. For tungsten-polymer composites, increasing the tungsten weight fraction increases acoustic impedance but decreases attenuation \cite{grewe1990acoustic}. Previous prototyping showed that a tungsten weight ratio of 80.8\% has a high enough impedance to provide damping, and a high enough absorption to attenuate internal backing layer reverberation to below the noise floor \cite{roberts2021}. A common backing layer was cast onto the rear electrodes of the PZT elements with a 24 mm thickness, and its rear face was given a scattering structure to further attenuate backing layer reverberation (Figure \ref{fig:assembly_summary_gallery}H-I).

A 400 $\mu$m layer of Aptflex F7 polyurethane (Precision Acoustics, Dorchester, UK) was added to the front face of the transducer, to provide electrical insulation. This material has an acoustic impedance of 1.5 MRayl, making the transmission coefficient at the coating-water boundary approximately 1.

The total material cost of the manufacture was £2k, comprising £906 for the PZT elements, £140 for 3D-printing filament, and £954 for off-the-shelf components, adhesives, and consumable materials. The total manufacture duration was 4 months of one person working full time, including the manufacture of custom tooling. Due to the large number of transducer modules and elements, most manufacture processes were typically performed in 8 separate batches, meaning that the time taken for 3D-printing parts was not a limiting factor, since this took place in parallel to the manual assembly steps.

\section{Nominal Acoustic Performance and IEV} \label{sect: inter_element_variation}

After the 16 array modules were manufactured, their nominal acoustic performance and IEV were characterised. All of the results are shown without normalisation, and are summarised in Table \ref{table: summary_stats}.

\subsection{Electrical Input Impedance} \label{sect: impedance}
Figure \ref{fig: electrical_impedance} shows the electrical input impedance of the transducer elements immersed in deionised water, measured after backing layer casting. The phase spectra has a peak at 1.22 MHz matching the series resonance of the in-air PZT plate, with a small phase angle indicating relatively weak damping. This could be because the PZT vibration is dominated by the coupling of lateral and thickness modes, which may not be effectively damped by the backing layer. Other UST transducers with a similar backing material had a larger resonance phase angle of -62.5$\degree$, and were damped laterally by epoxy \cite{angerer2020model}, but in this work the lateral damping was very weak, since the kerfs were filled with water during the impedance measurement (polyurethane for the final transducers), creating a large reflection coefficient at the PZT-kerf boundary.

The IEV in impedance magnitude and phase was low, with no defective channels. The $\pm$4.96 $\degree$ standard deviation in peak phase angle is very similar to the $\pm$5.7 $\degree$ standard deviation reported for 144 UST transducer elements manufactured using advanced equipment \cite{angerer2020semi}, with a smaller overall range. This demonstrates the reliability of the conductive-epoxy technique used to connect the PZT element electrodes to the PCB, and also demonstrates that the matching layers, PZT plates, and backing layers have uniform acoustic properties and dimensions. The ability to measure the electrical input impedance of the acoustic stack during manufacture is a useful interface, since it allows users to discard defective transducer modules as soon as possible or to collect data when making modifications, such as changing the PZT element dimensions. 

\subsection{Transmit Impulse Response} \label{sect: transmit_impulse_response}
\begin{figure}[!h]
\centerline{\includegraphics[width=0.5\columnwidth]{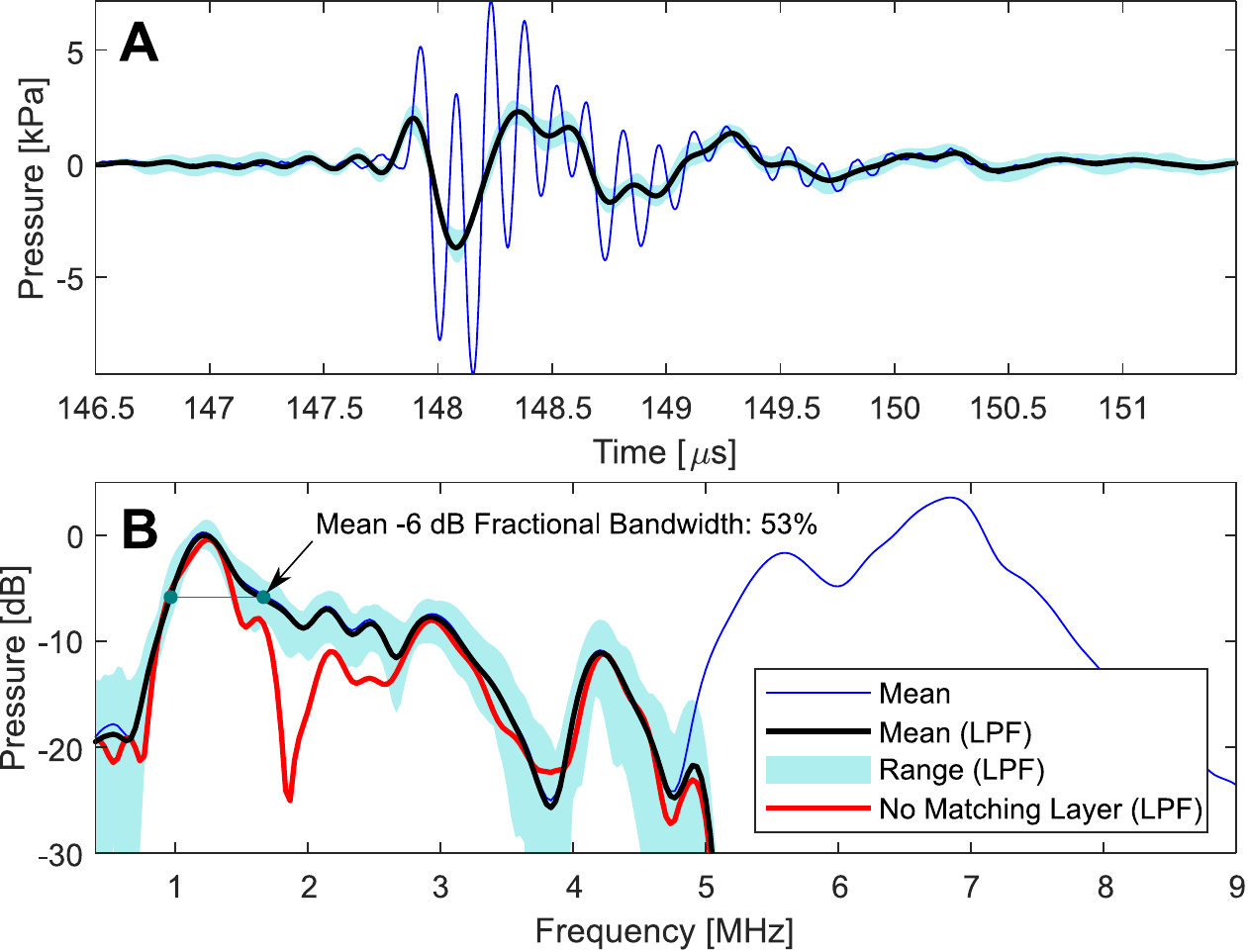}}
\caption{\textbf{A:} Transmit impulse response hydrophone signals, aligned in time. \textbf{B:} Amplitude spectra. Is shown before and after low pass filtering (LPF). The red line shows the mean amplitude spectra of N = 16 transducer elements without matching layers (displayed with the same dB reference).}
\label{fig:transmit_impulse_response_probes_AEFG}
\end{figure}

Figure \ref{fig:transmit_impulse_response_probes_AEFG} shows the mean transmit impulse response for 64 elements, measured using a calibrated 200 $\mu$m polyvinylidene fluoride needle hydrophone (Precision Acoustics, Dorchester, UK) after driving each element with a 80 ns pulse, which excited harmonics up to 6.8 MHz. The higher harmonics are not typically used for UST, so the impulse response was low pass filtered (cutoff 5 MHz) so that the waveform shape and IEV could be more easily visualised in the frequency range of interest.

The -6 dB and -12 dB fractional bandwidths are 53 \% and 175 \%, with pass-bands at 967 kHz - 1.67 MHz and 833 kHz - 3.23 MHz respectively. Figure \ref{fig:transmit_impulse_response_probes_AEFG} shows weak resonance features from the 1.22 MHz acoustic centre frequency up to 2.86 MHz, which correspond to the in-air PZT resonances shown in Figure \ref{fig: electrical_impedance}. The IEV in impulse response was low with no outliers, and only a small amplitude deviation at resonance of $\pm$6.9\%, which again demonstrates the consistency in the acoustic properties and dimensions of the matching layers, PZT plates, and backing layers. 

\begin{table}[t] 
\centering
\caption{Summary statistics for the open-UST transducer ring array.}
\label{table: summary_stats}
\begin{tabular}[t]{>{\raggedright}p{0.5\columnwidth}>{\raggedright}p{0.2\columnwidth}>{\raggedright\arraybackslash}p{0.15\columnwidth}} 
\toprule
Parameter&Mean Value&Standard Deviation\\
\midrule
\multicolumn{3}{l}{\textbf{Electrical input impedance (N = 128)}}\\
Resonance frequency&1.23 MHz&18 kHz\\
Phase at resonance&-8.58 $\degree$&4.96 $\degree$\\
Magnitude at resonance&1114 $\Omega$&121 $\Omega$\\
&&\\
\multicolumn{3}{l}{\textbf{Transmit impulse response (N = 64)}}\\
Resonance frequency&1.22 MHz&26 kHz\\
Amplitude deviation at resonance&--&6.9 \%\\
-6dB FBW (967 kHz - 1.67 MHz)&53 \%&12 \%\\
-12dB FBW (833 kHz - 3.23 MHz)&175 \% &32 \%\\
&&\\
\multicolumn{3}{l}{\textbf{Transmit impulse response, no matching layers (N = 16)}}\\
Resonance frequency&1.25 MHz&26 kHz\\
-6dB FBW (967 kHz - 1.43 MHz)&39 \%&3.2 \%\\
-12dB FBW (867 kHz - 1.73 MHz)&70 \% &2.4 \%\\
&&\\
\multicolumn{3}{l}{\textbf{Beam pattern (N = 48)}}\\
Elevational skew&0.457 $\degree$&0.207 $\degree$\\
Lateral skew&1.169 $\degree$&0.834 $\degree$\\
Elevational -6 dB beamwidth&16.3 mm&0.456 mm\\
Lateral -6 dB opening angle&55.4 $\degree$&2.96 $\degree$\\
&&\\
\multicolumn{3}{l}{\textbf{On-axis transmit-receive response (N = 256)}}\\
Resonance frequency&1.21 MHz&7.1 kHz\\
Amplitude deviation at resonance&--&7.9 \%\\
-6 dB FBW (1.08 MHz - 1.41 MHz)&29 \%&6.1 \%\\
-12 dB FBW (924 kHz - 1.58 MHz)&54 \% &4.9 \%\\
-40 dB  FBW (528 kHz - 2.61 MHz)&170 \%&3.0 \%\\
Signal to noise ratio&61.2 dB&1.2 dB\\
Arrival time&150.26 $\mu$s&0.070 $\mu$s\\  % range was 150.0800 to 150.4000
&&\\
\multicolumn{3}{l}{\textbf{Off-axis transmit-receive response (N = 256)}}\\
Receive cross talk&-37.1 dB&6.0 dB\\
\bottomrule
\end{tabular}
\end{table}

Figure \ref{fig:transmit_impulse_response_probes_AEFG} also shows the mean amplitude spectrum of a prototype 16-element module manufactured identically to the final transducers, but without matching layers. Comparing the two spectra shows that tuning the matching layer resonance to 2 MHz to boost the high frequency response was successful, since the -6 dB and -12 dB bandwidths increased from 39 \% and 70 \% to 53 \% and 175 \% respectively. At the 1.22 MHz centre frequency, the mean amplitude was 4.7 \% lower for the 16 elements measured without matching layers.

Users could tune the matching layer resonance by choosing a different thickness during manufacture, or the matching layers could be omitted altogether, which could reduce manufacture time by 2 months. Figure \ref{fig:transmit_impulse_response_probes_AEFG} shows that this may decrease image resolution due to the lower SNR above 1.4 MHz, but that the low frequency data required for FWI methods would be unaffected.

\subsection{Field Scans} \label{sect: field_scans} 

\begin{figure}[!h]
\centerline{\includegraphics[width=0.5\columnwidth]{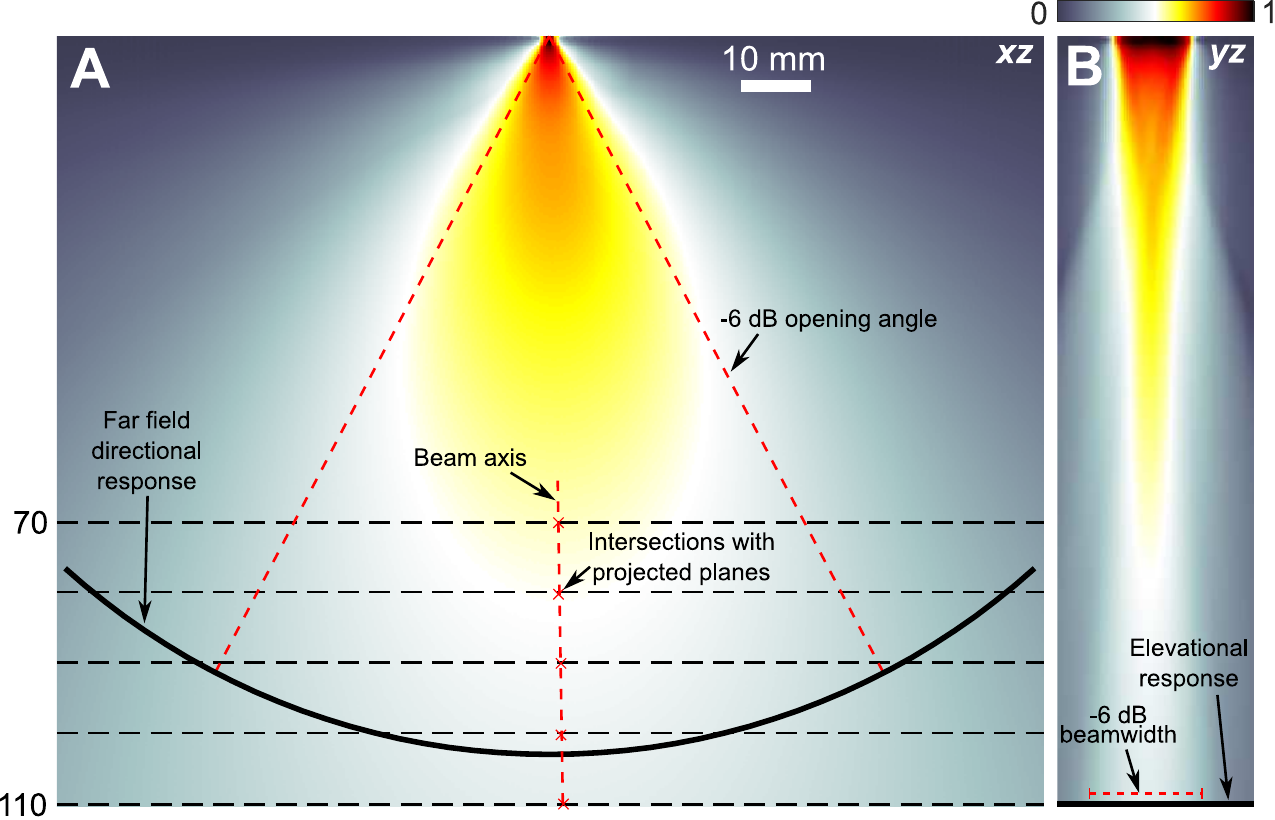}}
\caption{Peak pressure field for element 8, normalised and log compressed. \textbf{A}: Lateral plane and directional response profile calculated using 2D interpolation. \textbf{B:} Elevational plane and elevational response profile. The projected planes and the beam axis intersection points are also shown.}
\label{fig: beamplots}
\end{figure}

Figure \ref{fig: beamplots} shows the peak positive pressure field of a single element (channel 8). To reduce acquisition time, field scans were performed for 3 modules with all 16 elements driven simultaneously with a 1-cycle 1.4045 MHz 80 V tri-state pulse, which matches the driving conditions used for UST data acquisition. Hydrophone voltage signals were acquired over a 100.1 mm $\times$ 20.3 mm plane (0.35 mm step, 30.45 mm axial offset), within a time window including the entire pulse. The frequency dependent sensitivity of the hydrophone was deconvolved to obtain the pressure, and the measured field was backprojected to the source plane using the angular spectrum method \cite{zeng2008evaluation}. A mask was used to isolate the source field of each element, which was then re-projected forwards to 5 planes from z = 70 mm to z = 110 mm (for channel 8, the entire peak pressure field was projected for visualisation). The pressure amplitude field of each element $F(x, y, z, f)$ was calculated using a Fast Fourier Transform, and the beam axis was located at each axial $z$ position by calculating the weighted centroid $(x_{c}, y_{c}, f_{c})$ of a cross section through the field, ignoring values below -6dB (see Figure \ref{fig: beamplots}).

\subsection{Beam Axis Skew Angles} \label{sect: skew}
\begin{figure}[!h]
\centerline{\includegraphics[width=0.5\columnwidth]{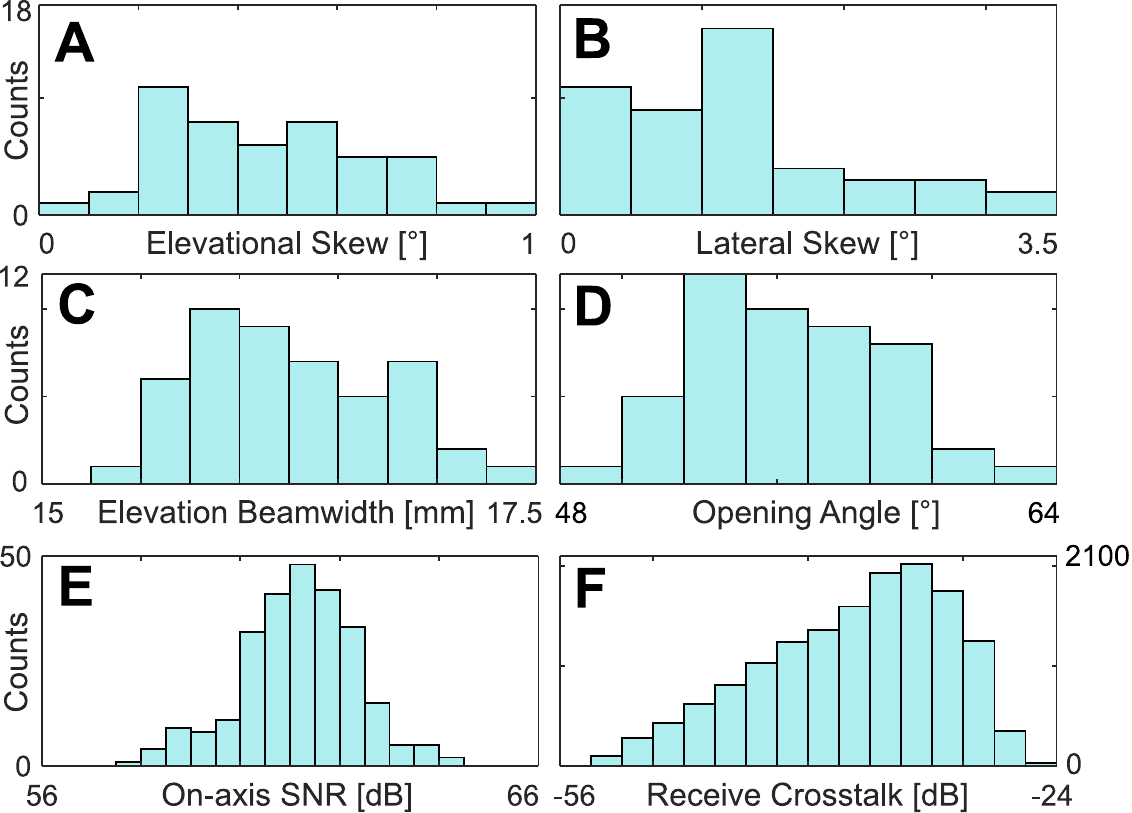}}
\caption{Histograms showing the distribution of elevational skew angle (\textbf{A}), lateral skew angle (\textbf{B}), elevational beamwidth (\textbf{C}), lateral opening angle (\textbf{D}), on-axis SNR (\textbf{E}) and receive cross talk (\textbf{F}).}
\label{fig:histograms}
\end{figure}

Figure \ref{fig:histograms} shows the distribution of elevational and lateral beam axis skew angles for 48 elements, calculated in the far field using linear fitting to the beam axis intersection points $(x_{c}, y_{c}, z)$ defined above. Transducer body misalignment relative to the hydrophone scan axes was estimated to be 0.115$\degree$ and 0.337$\degree$ in the elevation and lateral planes respectively, based on surface height data acquired from the transducer body, and by inspecting phase differences at the source plane. During UST data acquisition the transducer modules are mounted in the same fashion, and so the misalignment is also expected to be very small.

The small skew angles show that the PZT elements are well aligned relative to the transducer module, and that beam skew caused by non-uniformity in matching layer thickness is negligible. The lateral skew is larger than the elevational skew because the small element width makes alignment in this plane more sensitive to manufacture error. The beam axis skew angles are so small that they could be ignored during image reconstruction which simplifies the transducer modelling.

\subsection{Beamwidth, Opening Angle and Angle Dependent Frequency Response}

\begin{figure}[!h]
\centerline{\includegraphics[width=0.5\columnwidth]{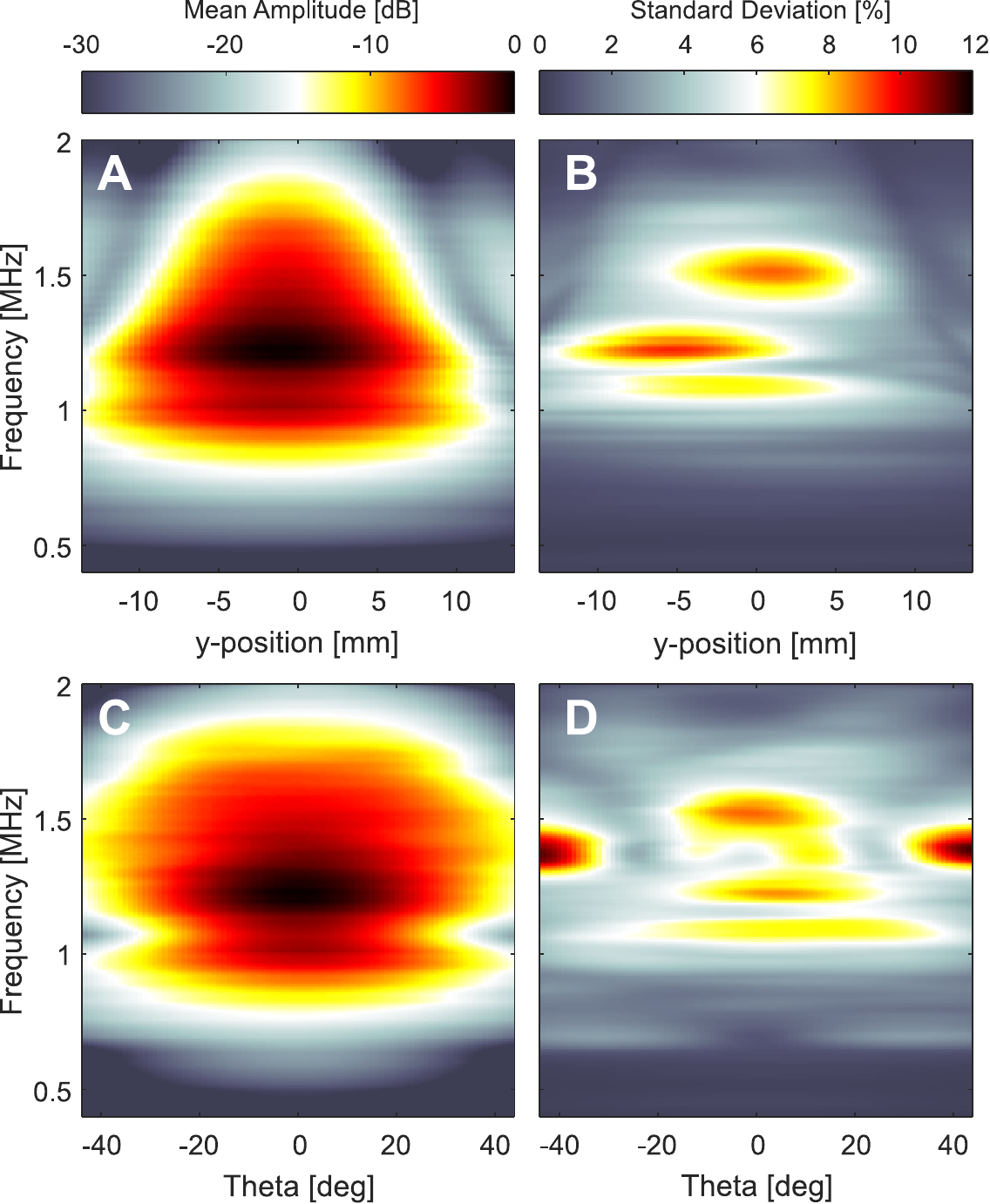}}
\caption{\textbf{A:} Mean elevational response. \textbf{B}: Standard deviation, relative to maximum of mean. \textbf{C}: Mean far field directional response. \textbf{D}: Standard deviation, relative to maximum of mean.}
\label{fig: spatial_impulse_response}
\end{figure}

Figure \ref{fig: spatial_impulse_response} shows the mean elevational response (amplitude as a function of elevation position $y$ and frequency $f$ at the centre of the ring array $z_c$ = 110 mm) and mean far field directional response of 48 elements, which were derived from the amplitude fields calculated previously (see Figure \ref{fig: beamplots}).

The ADR is smooth in both planes, suggesting that it could be easily incorporated into the forward model during image reconstruction. The standard deviation in the elevational and directional responses of the elements is low (maximum 9.5 \% and 11.5 \%), which is further evidence of the very small beam axis skew angles and the uniformity in the effective sizes of the elements. Figure \ref{fig:histograms}C shows the distribution of -6 dB elevation beamwidth (at the ring array centre) and the -6 dB opening angle, extracted at the centroid frequency $f_c$ from the elevational and directional responses respectively.

The nominal beamwidth of 16.3 mm at the ring array centre is relatively large, and may generate out-of-plane scattering artefacts if the images are reconstructed in 2D. However, in vivo data from a ring array with a 12 mm beamwidth has been successfully reconstructed using 2D FWI \cite{agudo20183d}, and for the same system a 3D forward model including the finite elevation beam has been shown to reduce out-of-plane artefacts \cite{sandhu20173d}. Therefore, the larger beamwidth of the open-UST system does not prevent its use as a research tool. The nominal beamwidth closely matches the predicted value of 16.4 mm at 1.22 MHz from the simulations in Section \ref{sect: pzt_selection}, demonstrating that the mean effective radiating length of the elements closely matches the ideal value of 10 mm. The nominal opening angle of 55.4$\degree$ is smaller than the predicted value of 95.1$\degree$ at 1.22 MHz from Figure \ref{fig:beamwith_opening_angle_AFP_simulations}A, which could be due to the strong lateral resonance modifying the radiating pressure, producing an effective source width larger than the physical extent of the elements \cite{delannoy1980anomalous}. However, this opening angle is suitable for imaging, since it is similar to the 43$\degree$ (-10 dB at 2.6 MHz) \cite{angerer2022single} opening angle reported for another UST system.

The low IEV in ADR is summarised by the small standard deviations in elevational beamwidth and lateral opening angle of 0.456 mm and 2.96$\degree$. This demonstrates that the radiating source pressure distribution was consistent between elements, meaning that the variation in matching layer geometry and acoustic properties was small.

\subsection{On-axis Transmit-Receive Response} \label{sect: watershot analysis}

\begin{figure}[!h]
\centerline{\includegraphics[width=0.5\columnwidth]{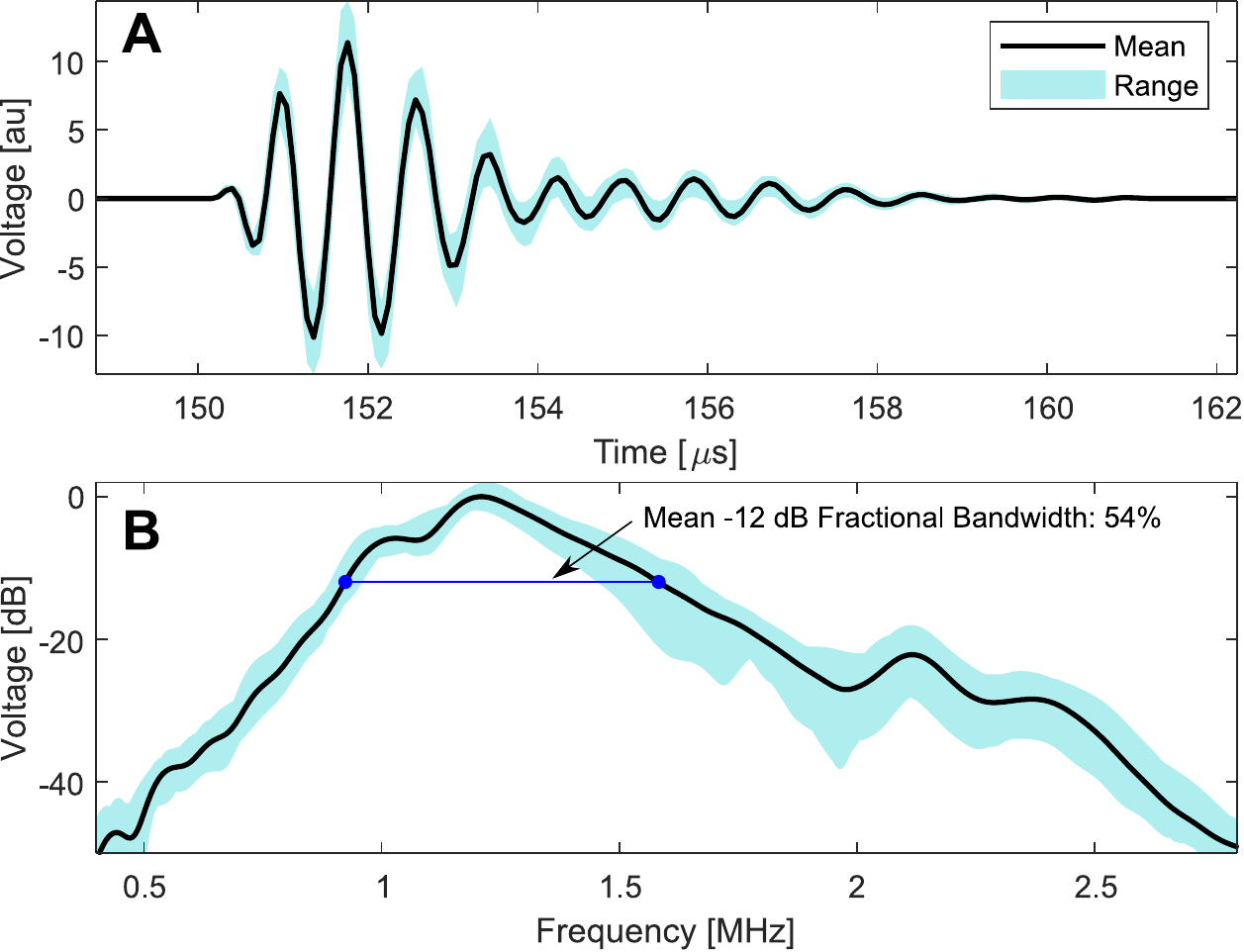}}
\caption{On-axis transmit-receive response. \textbf{A:} Measured voltages, aligned in time. \textbf{B:} Amplitude spectra. The mean and the entire measured range of the data are shown.}
\label{fig:tx_rx_on_axis_response}
\end{figure}

Figure \ref{fig:tx_rx_on_axis_response} shows the transmit-receive response for 256 on-axis transmit-receive element pairs, measured in a ring array configuration in deionised water. To acquire the watershot, each transmitter was driven with a 1-cycle 1.4045 MHz 80 V tri-state pulse, receiver data was measured on all other elements, and this was repeated for all transmitters. The transmit-receive bandwidth was 54 \% at -12 dB, and 170 \%  at -40 dB, with a centre frequency of 1.21 MHz. For FWI reconstruction, energy is required at low frequencies to generate a starting model. Excellent reconstructions have been achieved for data with a 766 kHz -40 dB cutoff frequency, starting in the 50 kHz - 500 kHz range \cite{bachmann2020source}. The open-UST -40 dB cutoff frequency is even lower at 528 kHz, and is therefore compatible with FWI methods.

The IEV in on-axis transmit-receive response of the transducer elements was low, with only a small amplitude deviation at resonance of 7.9 \%, which captures the combined uniformity in transmit pressure, receive sensitivity and beam axis alignment. Table \ref{table: summary_stats} shows a 0.07 $\mu$s standard deviation in arrival time for the on-axis signals, corresponding to a 104 $\mu$m deviation in acoustic path length or 8.4 \% of a cycle at 1.21 MHz. For comparison, the position errors for a UST bowl array manufactured with a 10 $\mu$m tolerance were between 300 $\mu$m and 1 mm \cite{tan2015newton}. This shows that the low cost techniques used for PZT element alignment are accurate. 

The distribution of on-axis SNR is shown in Figure \ref{fig:histograms}E, calculated using the first 7.9 $\mu$s of each received signal and 3.9 $\mu$s of noise, with a nominal value of 61.2 dB. This does not include the insertion loss due to breast tissue, which can be as high as 37 dB at 3.2 MHz \cite{sarno2022vivo}, or 12 dB at 1 MHz, assuming a linear frequency dependence. This would decrease the SNR to 49.2 dB, but this is still high and averaging could be used to improve SNR further.

\subsection{Directional Transmit-Receive Response}

\begin{figure}[!h]
\centerline{\includegraphics[width=0.5\columnwidth]{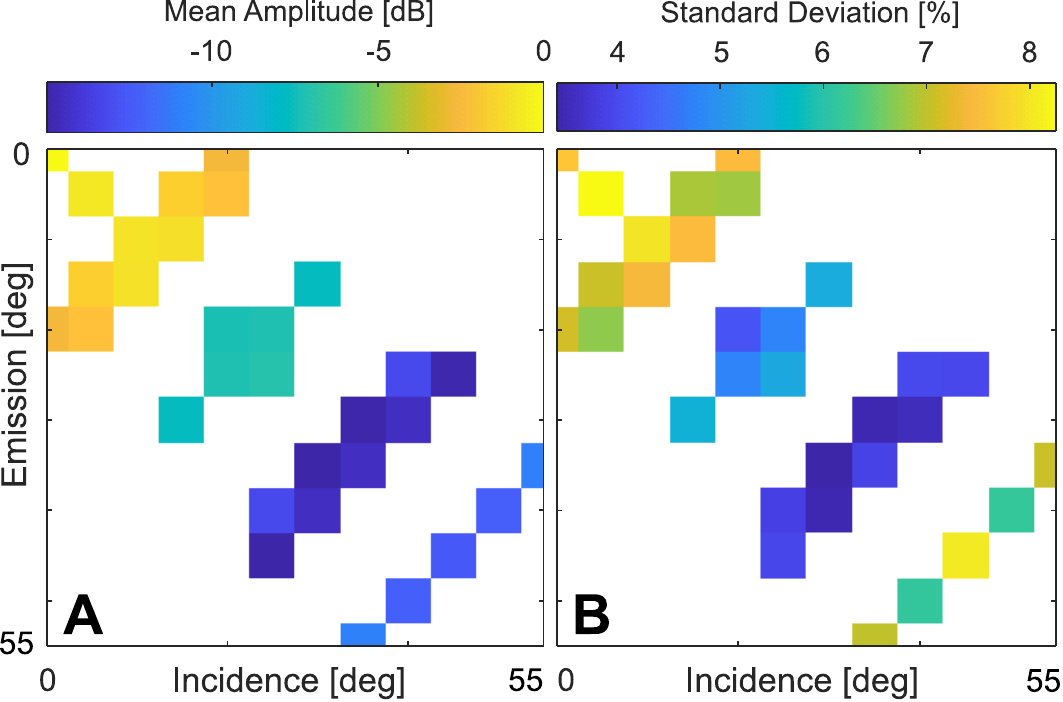}}
\caption{Transmit-receive directional response. \textbf{A:} Mean amplitude. \textbf{B:} Standard deviation, relative to maximum of mean.}
\label{fig: tx_rx_directional_response_mean_std}
\end{figure}

Figure \ref{fig: tx_rx_directional_response_mean_std} shows the off-axis transmit-receive response, defined as the peak value of the amplitude spectrum for the watershot dataset above, with the rays grouped into 5$\degree$ bins based on their emission and incidence angle.

Although data is not available for all bins, Figure \ref{fig: tx_rx_directional_response_mean_std} shows that the transmit-receive directional response is smooth, suggesting that the transducer elements could be modelled using an ideal rectangular source during image reconstruction. The SNR is also shown to be reduced by up to -14.9 dB when the emission and incidence angles are greater than 45$\degree$. The IEV in off-axis response was low, with a maximum standard deviation of 8.3\%, again showing that the beam axis skew angles are small, and that the effective radiating dimensions of the sources are uniform. 

\subsection{Receive Crosstalk}
\begin{figure}[!h]
\centerline{\includegraphics[width=0.5\columnwidth]{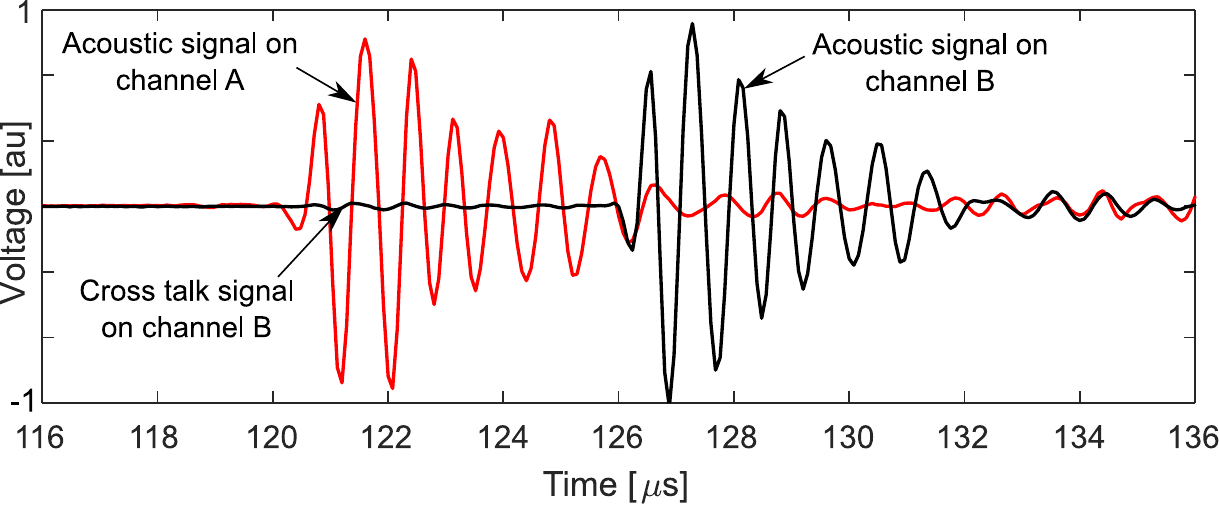}}
\caption{Example of receive crosstalk. A single aggressor channel is shown, but the crosstalk on the receptor channel is the superposition of the coupling to all other channels in the transducer module.}
\label{fig: Crosstalk}
\end{figure}

% Std  f_c over 36864 rays: 61447.716 Hz
Figure \ref{fig: Crosstalk} shows an example of receive cross talk between channels in the watershot data due to capacitive coupling in the bundled ribbon cable. The receive cross talk distribution is shown in Figure \ref{fig:histograms}F, defined as a power ratio between the cross talk and acoustic signal for each receive waveform. The mean cross talk was -37.1 dB, which did not affect the accuracy of the time of flight picking during the imaging experiment in Section \ref{sect: phantom_imaging}. The coupling could be reduced using microcoaxial cables or individually shielded twisted pairs, but these are expensive, less widely available and would increase manufacture time. Further work is required to assess the effect of the receive cross talk on FWI reconstructions. 

\section{Ultrasound Tomography Imaging Experiment} \label{sect: phantom_imaging}

\begin{figure}[!h]
\centerline{\includegraphics[width=0.5\columnwidth]{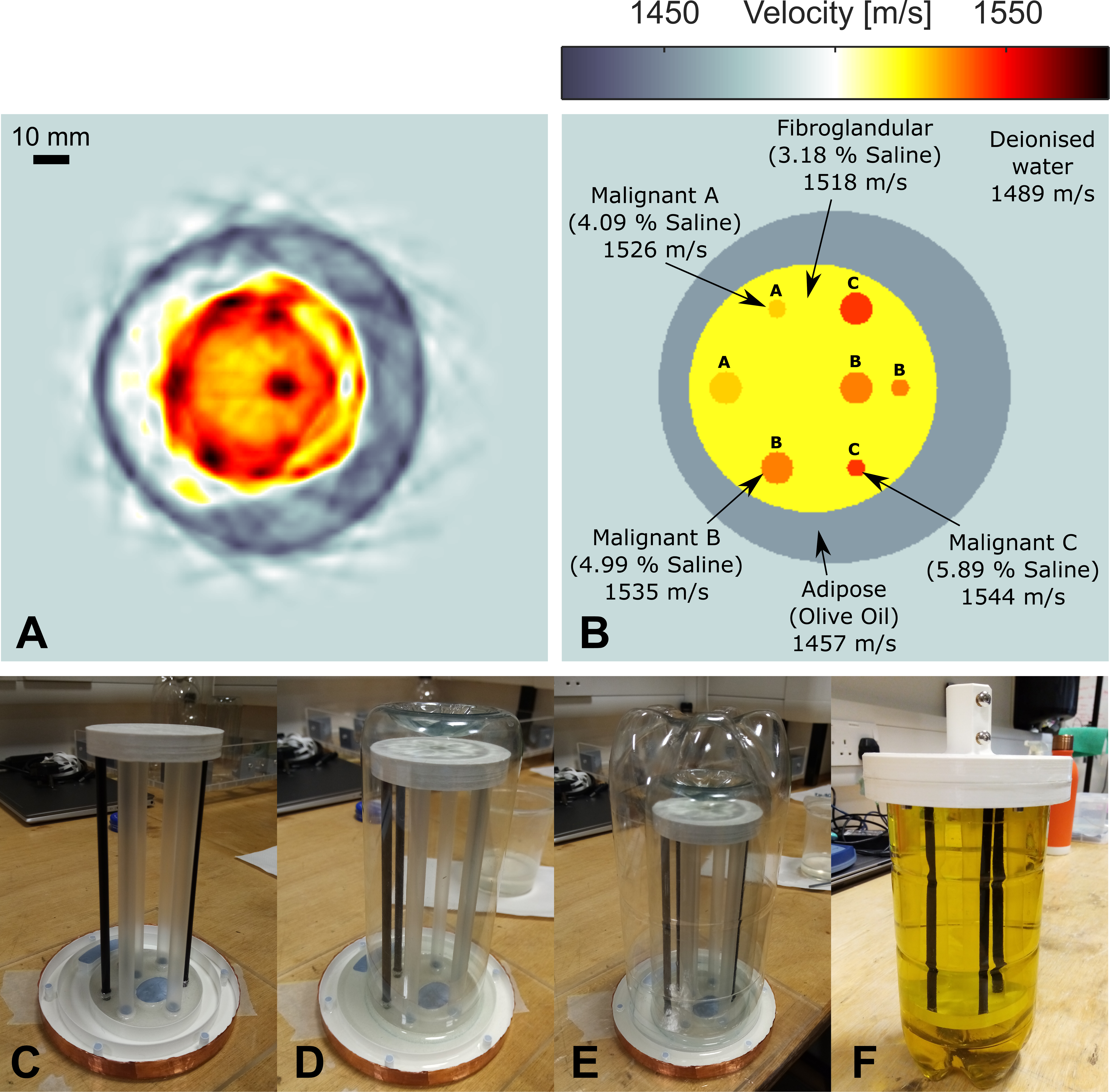}}
\caption{\textbf{A:} Reconstructed sound speed. \textbf{B:} True phantom sound speed map and tissue mimicking liquids. \textbf{C-F:} Phantom manufacture using PET bottles and straws.}
\label{fig:recon_phantom}
\end{figure}

A phantom UST experiment was performed to demonstrate suitability of the open-UST system for imaging research. The phantom was constructed to mimic the coronal plane of the breast, with a constant elevational cross section to reduce the out-of-plane errors arising from the finite elevation beamwidth of the transducer elements. Figure \ref{fig:recon_phantom}B-F shows the phantom manufacture, the tissue mimicking liquids and their sound speeds measured using through-transmission on homogeneous samples. Phantom and watershot UST datasets were acquired using the method described in Section \ref{sect: watershot analysis}.

Figure \ref{fig:recon_phantom}A shows the reconstructed sound speed, calculated using Kaczmarz's method of projections \cite{kak2001principles} to invert the relative time of flight data (this code has been made available on GitHub \cite{ustsart}). The adipose and fibroglandular regions, and all four of the 9 mm inclusions were resolved, but the three 5 mm inclusions and left fibroglandular boundary are distorted because the straight ray model does not capture refraction or diffraction. There is also a streaking artefact due to the relatively small number of elements. Nevertheless, this is a good proof of principle that the open-UST system is suitable for imaging.

\section{Discussion and Summary}
The total material cost of the 256-element transducer array including cables was £2k, which is very low. Users without access to a 3D-printer and vacuum chamber could purchase this equipment for $<$£5k. The cost could be reduced further by using a thin backing layer with a phase cancelling structure on the rear face \cite{ji2017phase} to reduce the required volumes of tungsten and epoxy. Also, the electromagnetic shielding could be omitted to reduce time and cost, and receive averaging could be used instead to reduce noise. The 4 month manufacture of the open-UST system is a short lead time for a transducer ring array, but adds staffing costs since the majority of the manufacture requires manual assembly. This could be addressed by omitting the matching layers, as discussed in Section \ref{sect: transmit_impulse_response}. For this work, a commercial 256-channel DAQ was used, but lower cost alternatives are available \cite{boni2018ultrasound}, or a multiplexer \cite{roman2018open} could be built to sequentially switch a pulser between each transmit channel and an oscilloscope between each receive channel, since this equipment is widely available.

The open-UST system has a similar cost and lead time to purchasing a pair of single element or clinical array transducers, and using a rotation stage to create a virtual array. However, these configurations have a significantly higher data acquisition time and mechanical complexity. Also, the open-UST system can be modified during the design phase, for example changing the ADR of the elements by adjusting their dimensions, which is not possible with off-the-shelf clinical probes, and would instead require expensive custom commercial arrays. 

Due to its open-source design, the functionality of the open-UST system could be extended by adding temperature measurement, on-board multiplexing, or electrical impedance matching to the interconnect PCB. The impulse response could be modified by adjusting the thickness and acoustic properties of the matching and backing layers. However, further work is needed to create a publicly available database of the acoustic properties of various metal filler/polymer composites, to reduce the upfront time spent on tuning the compositions to achieve the desired properties. The PZT element size could also be modified, but Section \ref{sect: acoustic_stack} demonstrated that it is not straightforward to predict the resonance behaviour of small PZT elements from their dimensions alone, without using finite element analysis. The open-UST system could also be a useful starting point for the rapid prototyping of low cost transducer arrays for applications outside of breast UST, for example in ultrasound therapy, rewarming, or industrial non destructive testing. 

The open-UST manufacture was designed to be accessible, without using specialist equipment. Tight manufacturing tolerances were achieved, but these depended heavily on calibrated offsets added to CAD models to compensate for systematic 3D-printing errors. Further work is needed to assess whether end users could replicate these results, without the experience gained during the prototyping phase.

The nominal bandwidth, beam pattern and SNR are similar to other UST arrays and are compatible with FWI methods, and so the open-UST system is suitable for UST imaging. The smooth ADR of the transducer elements could be modelled by representing the elements as ideal rectangular sources \cite{wise2019representing} during FWI reconstruction, with dimensions chosen to best match simulated and measured watershot datasets. This removes the need for an individual source definition based on additional hydrophone measurements, simplifying the calibration for the user. The electro-mechanical impulse response could also be derived from Figure \ref{fig:tx_rx_on_axis_response} using de-autoconvolution \cite{van2017impulse}.

Section \ref{sect: inter_element_variation} showed that the IEV in ADR was low, that the on-axis position errors were small enough to be calibrated using simple time of flight methods \cite{tan2015newton}, and that the beam axis skew angles were negligible. Therefore, users could model the transducers identically during image reconstruction, avoiding the need to characterise individual elements using hydrophone scans, which would add complexity. Only the ADR amplitude information was assessed in this paper, but since the image reconstruction was successful the IEV in the phase is also expected to be low. Further work is required to assess the reconstruction accuracy using FWI methods, in the case where the transducers are assumed to be identical.

The IEV in transmit-receive response was characterised for all of the elements in the array, and was similar to the IEV in the other characteristics. Therefore, the summary statistics in Table \ref{table: summary_stats} calculated for an array subset are likely to reflect the acoustic performance distribution of the entire array. The IEV in electrical impedance, fractional bandwidth, opening angle and element position were similar to other UST systems manufactured using advanced equipment. This demonstrates that the low cost techniques used for open-UST manufacturing framework also achieved high precision and low variation. 

This paper presented open-UST: a manufacturing framework for a low cost transducer ring array. The acoustic performance and inter-element variation were evaluated, and a phantom experiment was carried out demonstrating the suitability of open-UST for imaging research. A manufacture guide has been made available online \cite{openUST}.

%Bibliography
\bibliographystyle{unsrt}  
\bibliography{arxiv_open-UST}  

\begin{thebibliography}{10}

\bibitem{marmot2013benefits}
Michael~G Marmot, DG~Altman, DA~Cameron, JA~Dewar, SG~Thompson, and Maggie
  Wilcox.
\newblock {The benefits and harms of breast cancer screening: an independent
  review}.
\newblock {\em British journal of cancer}, 108(11):2205--2240, 2013.

\bibitem{wiskin2020full}
James Wiskin, Bilal Malik, David Borup, Nasser Pirshafiey, and John Klock.
\newblock {Full wave 3D inverse scattering transmission ultrasound tomography
  in the presence of high contrast}.
\newblock {\em Scientific Reports}, 10(1):1--14, 2020.

\bibitem{duric2020using}
Neb Duric, Mark Sak, Shaoqi Fan, Ruth~M Pfeiffer, Peter~J Littrup, Michael~S
  Simon, David~H Gorski, Haythem Ali, Kristen~S Purrington, Rachel~F Brem,
  et~al.
\newblock {Using whole breast ultrasound tomography to improve breast cancer
  risk assessment: A novel risk factor based on the quantitative tissue
  property of sound speed}.
\newblock {\em Journal of clinical medicine}, 9(2):367, 2020.

\bibitem{lucka2021high}
Felix Lucka, Mailyn P{\'e}rez-Liva, Bradley~E Treeby, and Ben~T Cox.
\newblock {High resolution 3D ultrasonic breast imaging by time-domain full
  waveform inversion}.
\newblock {\em Inverse Problems}, 38(2):025008, 2021.

\bibitem{bachmann2020source}
Etienne Bachmann and Jeroen Tromp.
\newblock {Source encoding for viscoacoustic ultrasound computed tomography}.
\newblock {\em The Journal of the Acoustical Society of America},
  147(5):3221--3235, 2020.

\bibitem{kim2014rapid}
Yohan Kim, Adam~D Maxwell, Timothy~L Hall, Zhen Xu, Kuang-Wei Lin, and
  Charles~A Cain.
\newblock {Rapid prototyping fabrication of focused ultrasound transducers}.
\newblock {\em IEEE transactions on ultrasonics, ferroelectrics, and frequency
  control}, 61(9):1559--1574, 2014.

\bibitem{yddal2015open}
Torstein Yddal, Sandy Cochran, Odd~Helge Gilja, Michiel Postema, and Spiros
  Kotopoulis.
\newblock {Open-source, high-throughput ultrasound treatment chamber}.
\newblock {\em Biomedical Engineering/Biomedizinische Technik}, 60(1):77--87,
  2015.

\bibitem{cueto2022stride}
Carlos Cueto, Oscar Bates, George Strong, Javier Cudeiro, Fabio Luporini,
  {\`O}scar~Calder{\'o}n Agudo, Gerard Gorman, Lluis Guasch, and Meng-Xing
  Tang.
\newblock {Stride: A flexible software platform for high-performance ultrasound
  computed tomography}.
\newblock {\em Computer Methods and Programs in Biomedicine}, 221:106855, 2022.

\bibitem{ruiter2018usct}
Nicole~V Ruiter, Michael Zapf, Torsten Hopp, Hartmut Gemmeke, Koen~WA van
  Dongen, Jorge Camacho, Joaqu{\'\i}n~L Herraiz, Mailyn~Perez Liva, and Jose~M
  Ud{\'\i}as.
\newblock {USCT reference data base: conclusions from the first SPIE USCT data
  challenge and future directions}.
\newblock In {\em Medical Imaging 2018: Ultrasonic Imaging and Tomography},
  volume 10580, pages 170--176. SPIE, 2018.

\bibitem{smith2022open}
Elliott Smith, Luzhen Nie, James McLaughlan, Harry Clegg, Thomas Carpenter,
  David Cowell, Stephen Evans, Alejandro~F Frangi, and Steven Freear.
\newblock {An Open Access Chamber Designed for the Acoustic Characterisation of
  Microbubbles}.
\newblock {\em Applied Sciences}, 12(4):1818, 2022.

\bibitem{marzo2017ultraino}
Asier Marzo, Tom Corkett, and Bruce~W Drinkwater.
\newblock {Ultraino: An open phased-array system for narrowband airborne
  ultrasound transmission}.
\newblock {\em IEEE transactions on ultrasonics, ferroelectrics, and frequency
  control}, 65(1):102--111, 2017.

\bibitem{openUST}
Morgan Roberts.
\newblock \url{morganjroberts.github.io/open-UST}, 2022.

\bibitem{cueto2021spatial}
Carlos Cueto, Lluis Guasch, Javier Cudeiro, Oscar~Calderon Agudo, Thomas
  Robins, Oscar Bates, George Strong, and Meng-Xing Tang.
\newblock {Spatial response identification enables robust experimental
  ultrasound computed tomography}.
\newblock {\em IEEE Transactions on Ultrasonics, Ferroelectrics, and Frequency
  Control}, 69(1):27--37, 2021.

\bibitem{roberts2021}
Morgan Roberts, Eleanor Martin, Michael Brown, Ben Cox, and Bradley Treeby.
\newblock {Transducer Module Development for an Open-Source Ultrasound
  Tomography System}.
\newblock In {\em 2021 IEEE International Ultrasonics Symposium (IUS)}, pages
  1--4, 2021.

\bibitem{roberts2022matching}
{M. Roberts, E. Martin, M. Brown, B. Cox, and B. Treeby}.
\newblock {Matching Layer Deposition for an Open-Source Ultrasound Tomography
  System: Inter-Element Variation in Frequency Response}.
\newblock In {\em 2022 IEEE International Ultrasonics Symposium (IUS)}, pages
  1--4. IEEE, 2022.

\bibitem{gemmeke20073d}
H~Gemmeke and NV~Ruiter.
\newblock {3D ultrasound computer tomography for medical imaging}.
\newblock {\em Nuclear Instruments and Methods in Physics Research Section A:
  Accelerators, Spectrometers, Detectors and Associated Equipment},
  580(2):1057--1065, 2007.

\bibitem{camacho2018multimodal}
J.~F.~Cruza J.~Camacho, N.~González-Salido, C.~Fritsch, M.~Pérez-Liva, J.L.
  Herraiz, and J.M. Udías.
\newblock {A Multi-Modal Ultrasound Breast Imaging System}.
\newblock In {\em Proceedings of the International Workshop on Medical
  Ultrasound Tomography: 1.- 3. Nov. 2017, Speyer, Germany}, pages 119--130.
  KIT Scientific Publishing, 2018.

\bibitem{zapf2022realization}
M~Zapf, T~Hopp, H~Gemmeke, M~Angerer, Z~Lu, O~Molchanova, N~Rashvand, R~Blanco,
  P~Steck, B~Leyrer, et~al.
\newblock {Realization of an pseudo-randomly sampled 3D USCT}.
\newblock In {\em Medical Imaging 2022: Ultrasonic Imaging and Tomography},
  volume 12038, pages 43--47. SPIE, 2022.

\bibitem{malik2018quantitative}
Bilal Malik, Robin Terry, James Wiskin, and Mark Lenox.
\newblock {Quantitative transmission ultrasound tomography: Imaging and
  performance characteristics}.
\newblock {\em Medical physics}, 45(7):3063--3075, 2018.

\bibitem{roy2013breast}
Olivier Roy, Steven Schmidt, Cuiping Li, Veerendra Allada, Erik West, David
  Kunz, and Neb Duric.
\newblock {Breast imaging using ultrasound tomography: From clinical
  requirements to system design}.
\newblock In {\em 2013 IEEE International Ultrasonics Symposium (IUS)}, pages
  1174--1177. IEEE, 2013.

\bibitem{CUDEIROBLANCO20221995}
Javier Cudeiro-Blanco, Carlos Cueto, Oscar Bates, George Strong, Tom Robins,
  Matthieu Toulemonde, Mike Warner, Meng-Xing Tang, Oscar~Calderón Agudo, and
  Lluis Guasch.
\newblock {Design and Construction of a Low-Frequency Ultrasound Acquisition
  Device for 2-D Brain Imaging Using Full-Waveform Inversion}.
\newblock {\em Ultrasound in Medicine \& Biology}, 48(10):1995--2008, 2022.

\bibitem{boni2018ultrasound}
Enrico Boni, CH~Alfred, Steven Freear, J{\o}rgen~Arendt Jensen, and Piero
  Tortoli.
\newblock {Ultrasound open platforms for next-generation imaging technique
  development}.
\newblock {\em IEEE transactions on ultrasonics, ferroelectrics, and frequency
  control}, 65(7):1078--1092, 2018.

\bibitem{agudo20183d}
O~Calderon Agudo, Lluis Guasch, Peter Huthwaite, and Michael Warner.
\newblock {3D imaging of the breast using full-waveform inversion}.
\newblock In {\em Proc. Int. Workshop Med. Ultrasound Tomogr.}, pages 99--110,
  2018.

\bibitem{huang2011characterization}
Shih-Ying Huang, John~M Boone, Kai Yang, Nathan~J Packard, Sarah~E McKenney,
  Nicolas~D Prionas, Karen~K Lindfors, and Martin~J Yaffe.
\newblock {The characterization of breast anatomical metrics using dedicated
  breast CT}.
\newblock {\em Medical physics}, 38(4):2180--2191, 2011.

\bibitem{apc_materials}
{Piezoelectric sensors: APC materials for Sensing Applications}, 2021.

\bibitem{treeby2018rapid}
Bradley~E Treeby, Jakub Budisky, Elliott~S Wise, Jiri Jaros, and BT~Cox.
\newblock {Rapid calculation of acoustic fields from arbitrary continuous-wave
  sources}.
\newblock {\em The Journal of the Acoustical Society of America},
  143(1):529--537, 2018.

\bibitem{angerer2022single}
Martin Angerer, Michael Zapf, Sylvia Gebhardt, and Nicole~V Ruiter.
\newblock {Single-PZT-Fiber Transducers for 3D Ultrasound Computed Tomography:
  Characterization and Modeling}.
\newblock In {\em 2022 IEEE International Ultrasonics Symposium (IUS)}, pages
  1--4. IEEE, 2022.

\bibitem{de1985vibration}
N~De~Jong, N~Bom, J~Souquet, and G~Faber.
\newblock {Vibration modes, matching layers and grating lobes}.
\newblock {\em Ultrasonics}, 23(4):176--182, 1985.

\bibitem{sandhu20173d}
Gursharan~Yash Sandhu, Erik West, Cuiping Li, Olivier Roy, and Neb Duric.
\newblock {3D frequency-domain ultrasound waveform tomography breast imaging}.
\newblock In {\em Medical Imaging 2017: Ultrasonic Imaging and Tomography},
  volume 10139, pages 56--69. SPIE, 2017.

\bibitem{angerer2020enhanced}
Martin Angerer, Michael Zapf, Sylvia Gebhardt, Holger Neubert, and Nicole~V
  Ruiter.
\newblock {Enhanced KLM model for single-fibre piezocomposite transducers}.
\newblock In {\em 2020 IEEE International Ultrasonics Symposium (IUS)}, pages
  1--4. IEEE, 2020.

\bibitem{roberts2019}
Morgan Roberts, Eleanor Martin, Ben Cox, and Bradley Treeby.
\newblock {Progress towards an open-source, low-cost ultrasound tomography
  research system}.
\newblock {\em Proceedings of International Workshop on Medical Ultrasound
  Tomography}, pages 53--66, 2019.

\bibitem{grewe1990acoustic}
Martha~G Grewe, TR~Gururaja, Thomas~R Shrout, and Robert~E Newnham.
\newblock {Acoustic properties of particle/polymer composites for ultrasonic
  transducer backing applications}.
\newblock {\em IEEE transactions on ultrasonics, ferroelectrics, and frequency
  control}, 37(6):506--514, 1990.

\bibitem{angerer2020model}
Martin Angerer, Michael Zapf, Benjamin Leyrer, and Nicole~V Ruiter.
\newblock {Model-guided manufacturing of transducer arrays based on
  single-fibre piezocomposites}.
\newblock {\em Applied Sciences}, 10(14):4927, 2020.

\bibitem{angerer2020semi}
{M. Angerer, M. Zapf, B. Leyrer, N. V. Ruiter}.
\newblock {Semi-automated packaging of transducer arrays for 3D ultrasound
  computer tomography}.
\newblock In {\em 2020 IEEE SENSORS}, pages 1--4. IEEE, 2020.

\bibitem{zeng2008evaluation}
Xiaozheng Zeng and Robert~J McGough.
\newblock {Evaluation of the angular spectrum approach for simulations of
  near-field pressures}.
\newblock {\em The Journal of the Acoustical Society of America},
  123(1):68--76, 2008.

\bibitem{delannoy1980anomalous}
B~Delannoy, C~Bruneel, F~Haine, and R~Torguet.
\newblock {Anomalous behavior in the radiation pattern of piezoelectric
  transducers induced by parasitic Lamb wave generation}.
\newblock {\em Journal of Applied Physics}, 51(7):3942--3948, 1980.

\bibitem{tan2015newton}
Wei~Yap Tan, Till Steiner, and Nicole~V Ruiter.
\newblock {Newton's method based self calibration for a 3D Ultrasound
  Tomography System}.
\newblock In {\em 2015 IEEE International Ultrasonics Symposium (IUS)}, pages
  1--4. IEEE, 2015.

\bibitem{sarno2022vivo}
Daniel Sarno, Christian Baker, Sian Curtis, Mark Hodnett, and Bajram Zeqiri.
\newblock {In Vivo Measurements of the Bulk Ultrasonic Attenuation Coefficient
  of Breast Tissue Using a Novel Phase-Insensitive Receiver}.
\newblock {\em IEEE Transactions on Ultrasonics, Ferroelectrics, and Frequency
  Control}, 69(10):2943--2954, 2022.

\bibitem{kak2001principles}
Avinash~C Kak and Malcolm Slaney.
\newblock {\em {Principles of computerized tomographic imaging}}.
\newblock SIAM, 2001.

\bibitem{ustsart}
Morgan Roberts.
\newblock \url{github.com/ucl-bug/ust-sart}, 2022.

\bibitem{ji2017phase}
Seon~Mi Ji, Jin~Ho Sung, Chan~Yuk Park, and Jong~Seob Jeong.
\newblock {Phase-canceled backing structure for lightweight ultrasonic
  transducer}.
\newblock {\em Sensors and Actuators A: Physical}, 260:161--168, 2017.

\bibitem{roman2018open}
Alex Roman, Parisa Dehghanzadeh, Vida Pashaei, Abhishek Basak, Swarup Bhunia,
  and Soumyajit Mandal.
\newblock {An open-source test-bench for autonomous ultrasound imaging}.
\newblock In {\em 2018 IEEE 61st International Midwest Symposium on Circuits
  and Systems (MWSCAS)}, pages 524--527. IEEE, 2018.

\bibitem{wise2019representing}
Elliott~S Wise, BT~Cox, Jiri Jaros, and Bradley~E Treeby.
\newblock {Representing arbitrary acoustic source and sensor distributions in
  Fourier collocation methods}.
\newblock {\em The Journal of the Acoustical Society of America},
  146(1):278--288, 2019.

\bibitem{van2017impulse}
Pim Van Der~Meulen, Pieter Kruizinga, Johannes~G Bosch, and Geert Leus.
\newblock {Impulse response estimation method for ultrasound arrays}.
\newblock In {\em 2017 IEEE International Ultrasonics Symposium (IUS)}, pages
  1--4. IEEE, 2017.

\end{thebibliography}

\end{document}